\documentclass[%
aps,
 reprint,
 amsmath,amssymb,
prper,
floatfix
]{revtex4-2}

\usepackage{graphicx}
\usepackage{dcolumn}
\usepackage{bm}

\newcommand{\attr}[1]{\textit{#1}}

\begin{document}


\title{Taking introductory physics in studio, lecture, or online format:\\ what difference does it make in subsequent courses, and for whom?}

\author{Gerd Kortemeyer}
 \email{kgerd@ethz.ch}
 \affiliation{%
 Educational Development and Technology, ETH Zurich, 8092 Zurich, Switzerland
}%
\altaffiliation[also at ]{Michigan State University, East Lansing, MI 48823, USA}

\author{Christine Kortemeyer}
 \affiliation{%
Lucerne University of Applied Sciences and Arts, 6002 Lucerne, Switzerland
}%

\author{Wolfgang Bauer}
 \email{bauerw@msu.edu}
 \affiliation{%
Department of Physics and Astronomy, Michigan State University, East Lansing, MI 48824, USA
}%

\date{\today}

\begin{abstract}
At large institutions of higher education, students frequently have a choice whether to attend the introductory physics sequence asynchronously online, on-site in a traditional lecture-setting, or in a reformed studio setting. In this study, we investigate how these different settings are correlated with measures of self-efficacy, interest in physics, and success in subsequent physics and engineering courses, which have the introductory physics sequence as prerequisites. As previous research indicates, some of these measures may depend on gender. We found that the course setting had no significant correlation with the grade in subsequent courses, but that studio-settings gave students the feeling of being better prepared, particularly for subsequent courses that included laboratory or recitation components. We also found that gender was  correlated with measures of interest in physics, where female students expressed significantly less interest in the subject, regardless of course setting.
\end{abstract}

\maketitle

\section{Introduction}
The influence of the course delivery modes on learning outcomes has been studied extensively across disciplines in general~\cite{russell1999no,cavanaugh2015,kortemeyer2023attending}, as well as for physics courses in particular~\cite{bergeler2021,kortemeyer22hybrid}; the topic received renewed attention due to the COVID-19 pandemic~\cite{jatmiko2021} While the general consensus appears to be that there is no significant difference regarding learning outcomes as measured by exams, there are hints of nuances: for example, students in online courses were found to be more likely to drop out, but those who persist are more likely to achieve higher grades~\cite{faulconer2018}. Also, certain face-to-face formats are more successful than others when it comes to conceptual understanding as measured by concept inventories, most notably the studio format~\cite{cummings1999,hoellwarth2005}. Unfortunately, passing courses with good grades as measured by exams can become the sole focus of students, particularly if interest in the subject is missing~\cite{lin}.

However, even artificial-intelligence agents could pass the assessments of calculus-based introductory physics courses~\cite{kortemeyer23ai}. Beyond the overt curriculum, course instructors frequently have additional objectives related to attitudes~\cite{brewe2009modeling}, expectations~\cite{mpex,sharma2013,zwickl2014}, curiosity~\cite{silverman1995self}, beliefs~\cite{madsen2015physics},  ``thinking like a physicist''~\cite{heuvelen}, and thinking of themselves as a physicist; the latter touches on issues of identity~\cite{irving2013} and self-efficacy~\cite{sawtelle2012exploring}. These epistemological factors may eventually influence learning success in the overt curriculum~\cite{may02,lising05e}.

Cwik and Singh studied these beliefs about physics and about students' physics learning longitudinally in a two-semester introductory physics course sequence where women are not underrepresented~\cite{cwik2022}, in particular self-efficacy, identity, and interest in physics.  They found significant differences between male and female learners in self-perceived competence, seeing themselves as a ``physics person,'' and  their expressed interest in physics. 

In this longitudinal study, we surveyed the students at the end of engineering or advanced physics courses for which the introductory calculus-based sequence are prerequisites. Students were free to choose online asynchronous, face-to-face lecture, or studio-based courses. Our main research question is whether the mode of instruction of the introductory sequence influences success in future courses, as well as retained self-efficacy and interest in physics, with a particular focus on gender.

\section{Setting}
Michigan State University is a public, large-enrollment ($>50,000$ students) R-1 university. Almost 78\% of the undergraduate population are from Michigan. About 47\% of the students identify as male, 53\% as female.

We considered three types of calculus-based introductory physics courses, namely asynchronous online, lecture-based, and studio-based offerings. Students were free to choose which type of course they enrolled into (even though, their choice may have been limited by external factors~\cite{kortemeyer2023attending}). The online courses were taught asynchronously using a variety of multimedia components~\cite{kortemeyer2014onl}. The lecture-based courses were partially flipped, but included traditional in-person lectures~\cite{kortemeyer22hybrid}. Finally, the studio-based courses were taught using the Projects and Practices in Physics (P-Cubed) pedagogy, which is a highly interactive, community-of-practice (COP) approach~\cite{beichner2007,irving2020}.

As subsequent courses we considered civil, mechanical and electrical engineering courses, as well as more advanced physics courses, for which the calculus-based introductory sequence is a prerequisite. Some of these had associated laboratories or recitations, while others had not.

\section{Methodology}
An online survey was anonymously distributed at the end of physics and engineering courses for which the calculus-based physics courses are prerequisites.
Table~\ref{tab:variables} shows the surveyed variables. We adopted the physics self-efficacy and interest variables from the study by Cwik and Singh~\cite{cwik2022}, and in addition, students were asked which introductory physics courses they had previously taken and what their grades in the altogether three courses are.

\begin{table*}
\caption{\label{tab:variables}Survey variables used in this study.}
\begin{ruledtabular}
\begin{tabular}{llp{8.8cm}}
Label			&Values							&Survey Prompt or Definition\\\hline
\attr{Concepts}		&NO!$=0$ -- YES!$=3$				&``I understand concepts I have studied in physics''\\
\attr{Course1}		&transfer, online, lecture, studio			&Type of first-semester introductory course\\
\attr{Course2}		&transfer, online, lecture, studio			&Type of second-semester introductory course\\
\attr{Gender}		&male$=0$, diverse$=0.5$, female$=1$	&Gender\\	 
\attr{Grade1st}		&worst$=0$ -- best$=4$				&Grade in first-semester introductory course\\
\attr{Grade2nd}		&worst$=0$ -- best$=4$				&Grade in second-semester introductory course\\
\attr{Grade3rd}		&worst$=0$ -- best$=4$				&Grade in third, subsequent course\\
\attr{HelpLabRec}	&NO!$=0$ -- YES!$=3$; N/A			&``I was able to help my classmates with physics in the laboratory or recitation attached to this course''\\
\attr{InterPhys}		&very~boring$=0$ -- very~interesting$=3$&``In general, I find physics''\\
\attr{KnowPhys}	&NO!$=0$ -- YES!$=3$				&``I want to know everything I can about physics''\\
\attr{Lecture}		&0 -- 2							&Number of lecture-based introductory courses taken\\
\attr{OcSetbcks}	&NO!$=0$ -- YES!$=3$				&``If I encounter a setback in a physics exam, I can overcome it''\\
\attr{Online}		&0 -- 2							&Number of online introductory courses taken\\
\attr{Prep}			&NO!$=0$ -- YES!$=3$				&``Did you feel that your previous physics courses prepared you well for this course?''\\
\attr{PrepProbSolve}	&not~at~all$=0$ -- very$=3$			&``How helpful were your previous physics courses for this course in terms of methods and problem solving?''\\
\attr{PrepTopics}	&not~at~all$=0$ -- very$=3$			&``How helpful were your previous physics courses for this course in terms of topic coverage?''\\
\attr{RecDiscs}		&NO!$=0$ -- YES!$=3$				&``I am curious about recent discoveries in physics''\\
\attr{Studio}		&0 -- 2							&Number of studio-based introductory courses taken\\
\attr{TestIfStudy}	&NO!$=0$ -- YES!$=3$				&``If I study, I will do well on a physics test''\\
\attr{WonderPhys}	&NO!$=0$ -- YES!$=3$				&``I wonder about how physics works''
\end{tabular}
\end{ruledtabular}
\end{table*}

For the subsequent third courses (denoted \attr{AnyCourse3}) we distinguished between those that had laboratory or recitation components (denoted \attr{LabRecCourse3}) and those that did not (denoted \attr{NoLabRecCourse3}). We also separately considered the advanced physics courses (denoted \attr{PhysicsCourse3}). These distinctions allowed us to separately evaluate the relationships between the variables in Table~\ref{tab:variables} within the subsamples of students having enrolled in these different classes of subsequent courses.

\section{Results}
\subsection{Demographics}
A total of 107~survey responses were received, of which 104~were completely filled out. Of their respondents, 65~identified as male, 38~as female, and 1~as diverse (women were underrepresented by almost $2:1$, compared to the approximately $1:1$ distribution in the overall student population). Eight respondents had transfer credit for only the first-semester course, and another four for both semesters; these courses were not counted in \attr{Online}, \attr{Lecture}, or \attr{Studio}, since their instructional mode was unknown.

Female students achieved slightly higher average grades than their male counterparts in all three semesters: with $4.0$ being the best grade, female students had averages grade of $3.7\pm0.5$, $3.7\pm0.4$, and $3.7\pm0.5$ for the first, second, and third course, respectively, compared to  $3.5\pm0.6$, $3.5\pm0.7$, and $3.4\pm0.7$ for the male students.

Table~\ref{tab:classes} shows the introductory physics courses taken by the respondents. Of the 104~respondents, 74~subsequently attended a third-semester physics or engineering course that included laboratory or recitation components (\attr{LabRecCourse3}), and 30~students took a subsequent physics course (\attr{PhysicsCourse3}).

\begin{table}
\caption{\label{tab:classes}Types of courses taken by the respondents.}
\begin{ruledtabular}
\begin{tabular}{lll}
Course Type		&Only one Semester	&Both Semesters\\
		\hline
Unknown	&11				&6\\
Online	&21				&6\\
Lecture	&47				&29\\
Studio	&33				&7
\end{tabular}
\end{ruledtabular}
\end{table}

\subsection{Correlations of Attributes}
\begin{figure*}
\begin{center}
\includegraphics[width=0.9\textwidth]{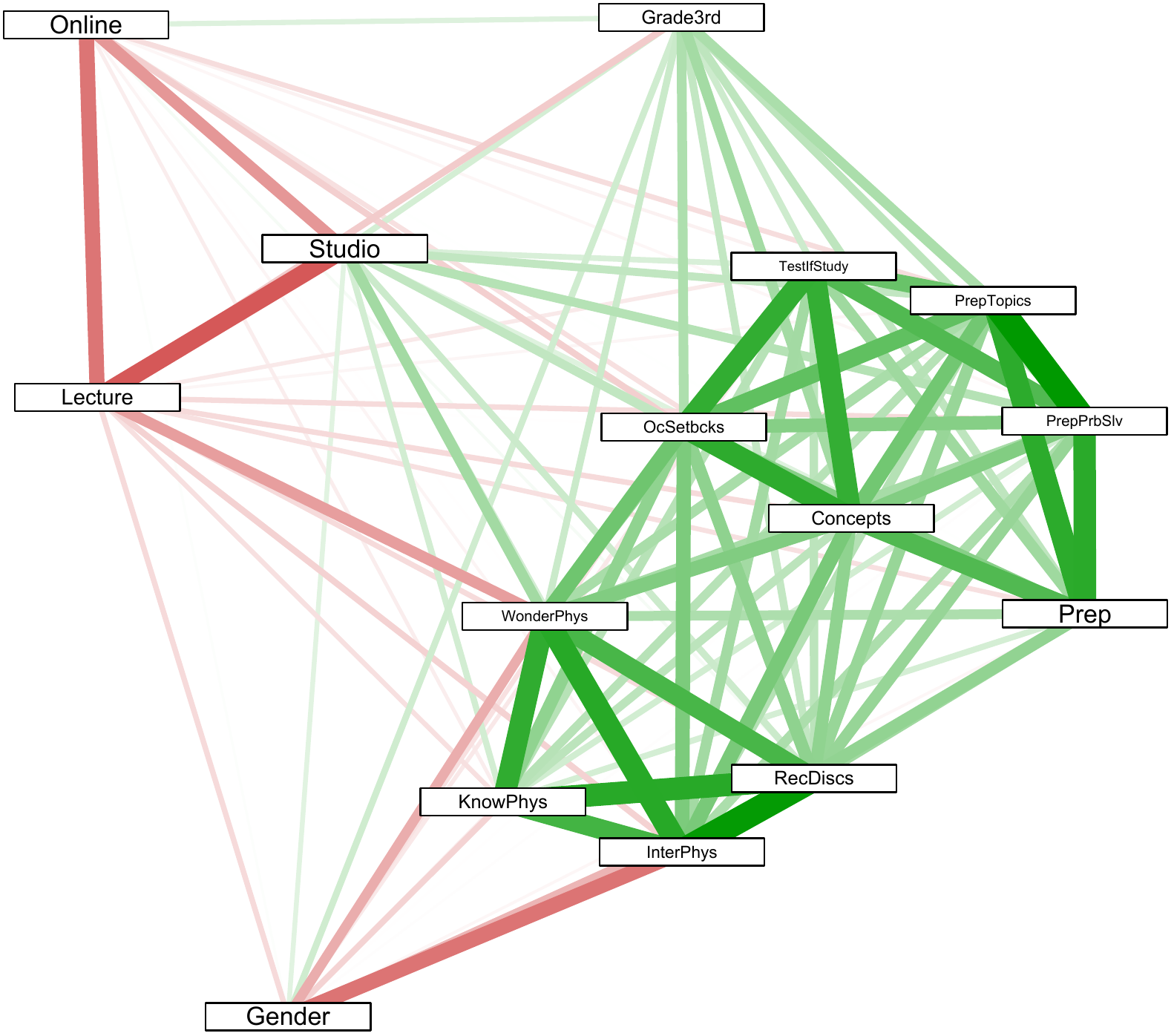}
\end{center}
\caption{Fruchterman-Reingold representation~\cite{fruchterman1991,qgraph} of the correlations between a subset of the survey variables (Tab.~\ref{tab:variables}, excluding \attr{HelpLabRec}) for any subsequent courses (\attr{AnyCourse3}, 104~respondents).\label{fig:correls}}
\end{figure*}

\begin{figure*}
\begin{center}
\includegraphics[width=0.9\textwidth]{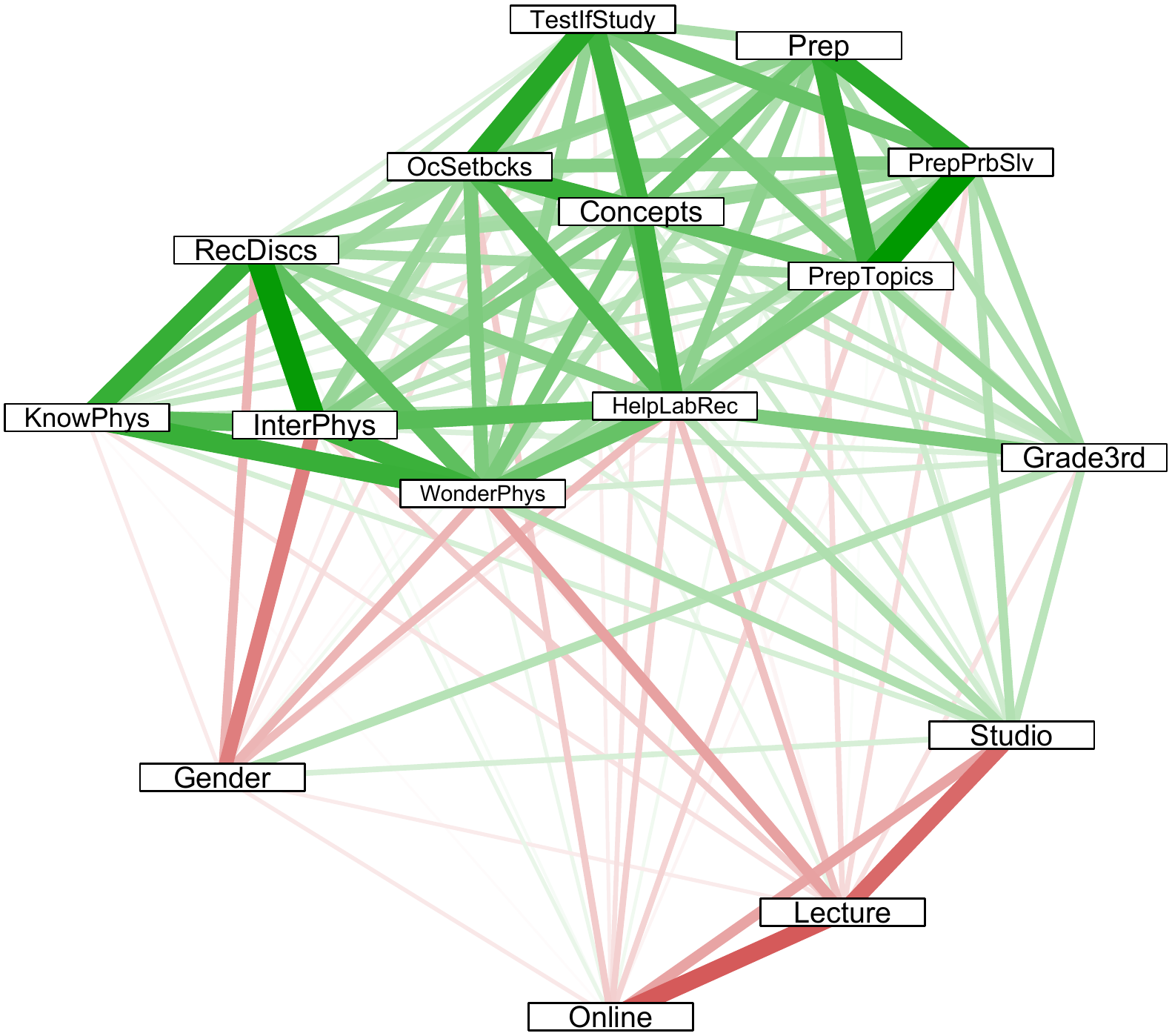}
\end{center}
\caption{Fruchterman-Reingold representation~\cite{fruchterman1991,qgraph} of the correlations between a subset of the survey variables (Tab.~\ref{tab:variables}, including \attr{HelpLabRec}) for subsequent courses that had laboratory or recitations components (\attr{LabRecCourse3}, 74~respondents).\label{fig:correlslab}}
\end{figure*}

Figures~\ref{fig:correls}--\ref{fig:correlslab} show the correlations between the variables in Tab.~\ref{tab:variables} as a force-directed Fruchterman-Reingold graph~\cite{fruchterman1991,qgraph,rproject}. The vertices denote the variables from Table~\ref{tab:variables}. Green edges denote positive correlations, while red edges denote negative ones; the thickness and saturation of these edges denotes the correlation strength. The distance between the vertices is determined by three sets of forces: a general repulsive force between the vertices, a central force to keep the vertices from drifting apart, and pairwise attractive forces that increase with the absolute value of the correlation~\cite{kortemeyer2022virtual}. Thus, mutually closely correlated vertices tend to cluster together, while unrelated vertices are farther apart. The rotation and handedness of the graphs is random.
Figure~\ref{fig:correls} shows this for \attr{AnyCourse3}, while Figures~\ref{fig:correlslab} only considers the subsample \attr{RecLabCourse3} and includes the \attr{HelpLabRec} vertex. 

While Figures~\ref{fig:correls} and~\ref{fig:correlslab} reflect trends in our sample and subsamples, not all of these correlations are significant, and not all of them are interesting. The negative correlations between the  introductory course types are trivial, and the positive correlations between several of the attributes are hardly surprising. Table~\ref{tab:sigcorrel} lists the non-trivial significant correlations between \attr{Online}, \attr{Lecture}, \attr{Studio}, and \attr{Gender} and other attributes~\cite{hmisc}.

\begin{table*}
\caption{\label{tab:sigcorrel}Significant correlations between attributes in Table~\ref{tab:variables}.}
\begin{ruledtabular}
\begin{tabular}{lllrl}
Attribute~1	&Attribute~2	&Sample/Subsample	&$r$-coefficient		&$p$-value\\\hline
\attr{Lecture}	&\attr{WonderPhys}	&\attr{AnyCourse3}	&$-0.27$	&$0.005$\\
\attr{Lecture}	&\attr{WonderPhys}	&\attr{LabRegCourse3}	&$-0.28$	&$0.02$\\
\attr{Lecture}	&\attr{Gender}	&\attr{PhysicsCourse3}	&$-0.49$	&$0.006$\\
\attr{Studio}	&\attr{OcSetbcks}	&\attr{NoLabRecCourse3}	&$0.36$	&$0.05$\\
\attr{Studio}	&\attr{WonderPhys}	&\attr{AnyCourse3}	&$0.26$	&$0.008$\\
\attr{Studio}	&\attr{WonderPhys}	&\attr{LabRegCourse3}	&$0.23$	&$0.04$\\
\attr{Studio}	&\attr{PrepPrbSlv}	&\attr{AnyCourse3}	&$0.21$	&$0.03$\\
\attr{Studio}	&\attr{PrepPrbSlv}	&\attr{LabRegCourse3}	&$0.23$	&$0.05$\\
\attr{Studio}	&\attr{HelpLabRec}	&\attr{LabRegCourse3}	&$0.24$	&$0.04$\\
\attr{Gender}	&\attr{WonderPhys}	&\attr{AnyCourse3}	&$-0.23$	&$0.02$\\
\attr{Gender}	&\attr{RecDiscs}	&\attr{AnyCourse3}	&$-0.21$	&$0.03$\\
\attr{Gender}	&\attr{InterPhys}	&\attr{AnyCourse3}	&$-0.39$	&$0.00004$\\
\attr{Gender}	&\attr{InterPhys}	&\attr{NoLabRecCourse3}	&$-0.44$	&$0.01$\\
\attr{Gender}	&\attr{InterPhys}	&\attr{LabRegCourse3}	&$-0.38$	&$0.0009$\\
\attr{Gender}	&\attr{InterPhys}	&\attr{PhysicsCourse3}	&$-0.46$	&$0.01$\\
\end{tabular}
\end{ruledtabular}
\end{table*}

It turns out that none of the variables are significantly correlated with course grades, and none were found for having attended the online courses. Instead, significant correlations emerged between a number of other variables: lecture courses were less likely selected by students expressing curiosity about physics (\attr{WonderPhys}), and they were also less likely selected by female students who later moved on to more advanced physics courses. Studio physics on the other hand were positively correlated with curiosity about physics and the feeling of preparedness for future courses, particularly in the area of problem solving. Female students generally expressed significantly less interest in physics ($p<0.0001$).

\subsection{Grade in Third Course based on Attributes}
Notably, in Table~\ref{tab:sigcorrel}, for neither the sample nor any of the subsamples, significant correlations ($p<0.05$) could be found between the introductory course types (\attr{Online}, \attr{Lecture}, or \attr{Studio}) and the grade in the subsequent course (\attr{Grade3rd}); this was confirmed by a multiple linear regression, which also showed no significance of the type of course for the grade in the subsequent course. In fact, in multiple linear regression only resulted in any significant relationships for the subsample \attr{LabRegCourse3}, see Table~\ref{tab:multreg} ($R^2=0.38$), but neither for the full sample nor any other subsamples.

\begin{table}
\caption{\label{tab:multreg}Multiple linear regression for the prediction of the grade in the third course, based on the attributes in Table~\ref{tab:variables}, for the subsample \attr{LabRegCourse3}  ($R^2=0.38$).}
\begin{ruledtabular}
\begin{tabular}{lrrrl}
Attribute	&$r$-coefficient	&Std.~Err.	&$t$-value	&$p$-value\\\hline
\attr{Online}    &   $0.24$   &$0.16$&$1.53 $&$0.132    $\\
\attr{Lecture}  &    $0.17$   &$0.15   $&$1.15$&$ 0.254  $\\  
\attr{Studio}     & $0.21$    &$0.17   $&$1.25 $&$0.218    $\\
\attr{Concepts} &   $ -0.12$    &$0.14 $&$ -0.87 $&$0.390  $\\  
\attr{TestIfStudy} & $0.15$    &$0.13 $&$  1.17 $&$0.246  $\\  
\attr{OcSetbcks}  & $-0.13$    &$0.12 $&$ -1.11 $&$0.271   $\\ 
\attr{WonderPhys} &  $0.04$    &$0.13  $&$ 0.35$&$ 0.730   $\\ 
\attr{KnowPhys}   &  $0.02$   &$0.12$&$   0.13 $&$0.894   $\\ 
\attr{RecDiscs}    & $0.24$    &$0.14 $&$  1.77 $&$0.0813  $\\ 
\attr{InterPhys}  & $-0.49$    &$0.21 $&$ -2.34$&$ 0.0226$\\ 
\attr{Prep}       &  $0.09$    &$0.11 $&$  0.84 $&$0.403    $\\
\attr{PrepPrbSlv} & $-0.10$    &$0.12 $&$ -0.79$&$ 0.430  $\\  
\attr{PrepTopics}  & $0.10$    &$0.12 $&$  0.83$&$ 0.409   $\\ 
\attr{Gender}     &  $0.33$    &$0.15 $&$  2.15 $&$0.0359  $\\
\attr{HelpLabRec} &  $0.38$    &$0.10  $&$ 3.76$&$ 0.000395$\\\\
Intercept&$ 2.92$&$    0.46$&$   6.29$&$ 4.59\cdot10^{-8}$
\end{tabular}
\end{ruledtabular}
\end{table}

For the subsample \attr{LabRecCourse}, feeling able to help in laboratory and recitation settings (\attr{HelpLabRec}) has the most significant impact on the grade in the third course, followed by interest in physics (\attr{InterPhys}) and female students doing better (\attr{Gender}). Even within this subsample, the other attributes had no significant correlation, including the course type. 

\subsection{Correlation between Grades}
\begin{figure*}
\begin{center}
\includegraphics[width=0.33\textwidth]{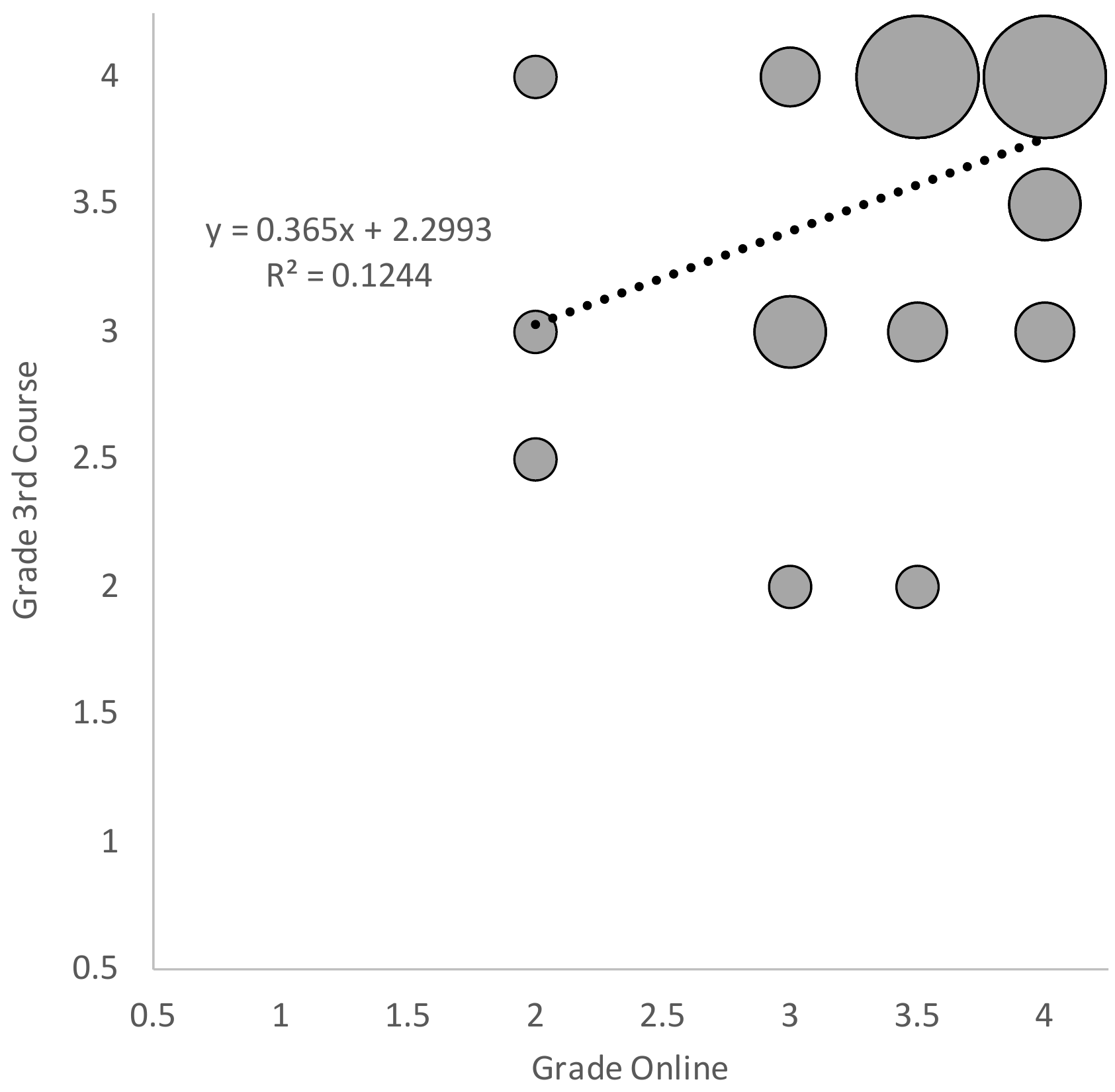}\includegraphics[width=0.33\textwidth]{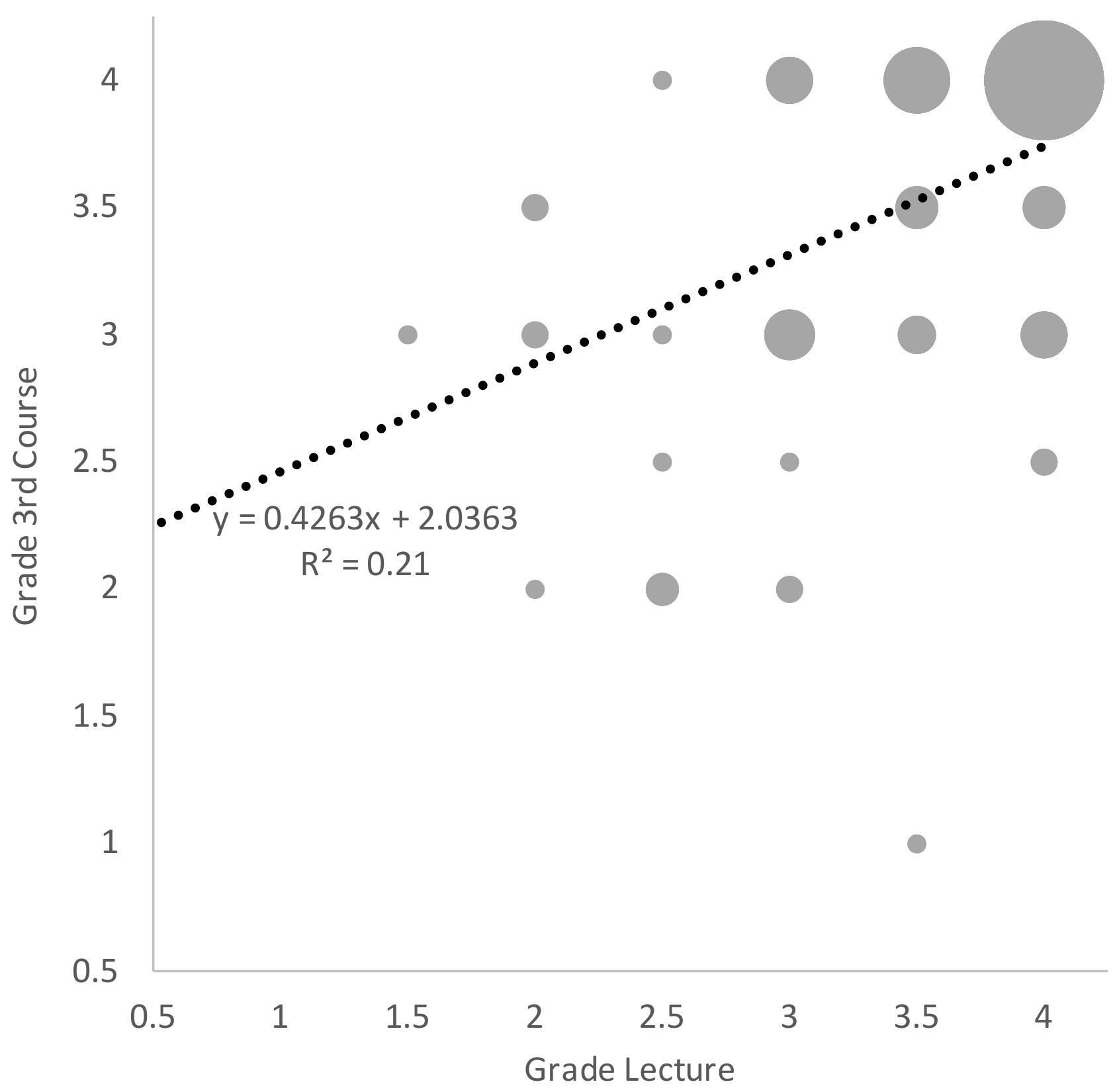}\includegraphics[width=0.33\textwidth]{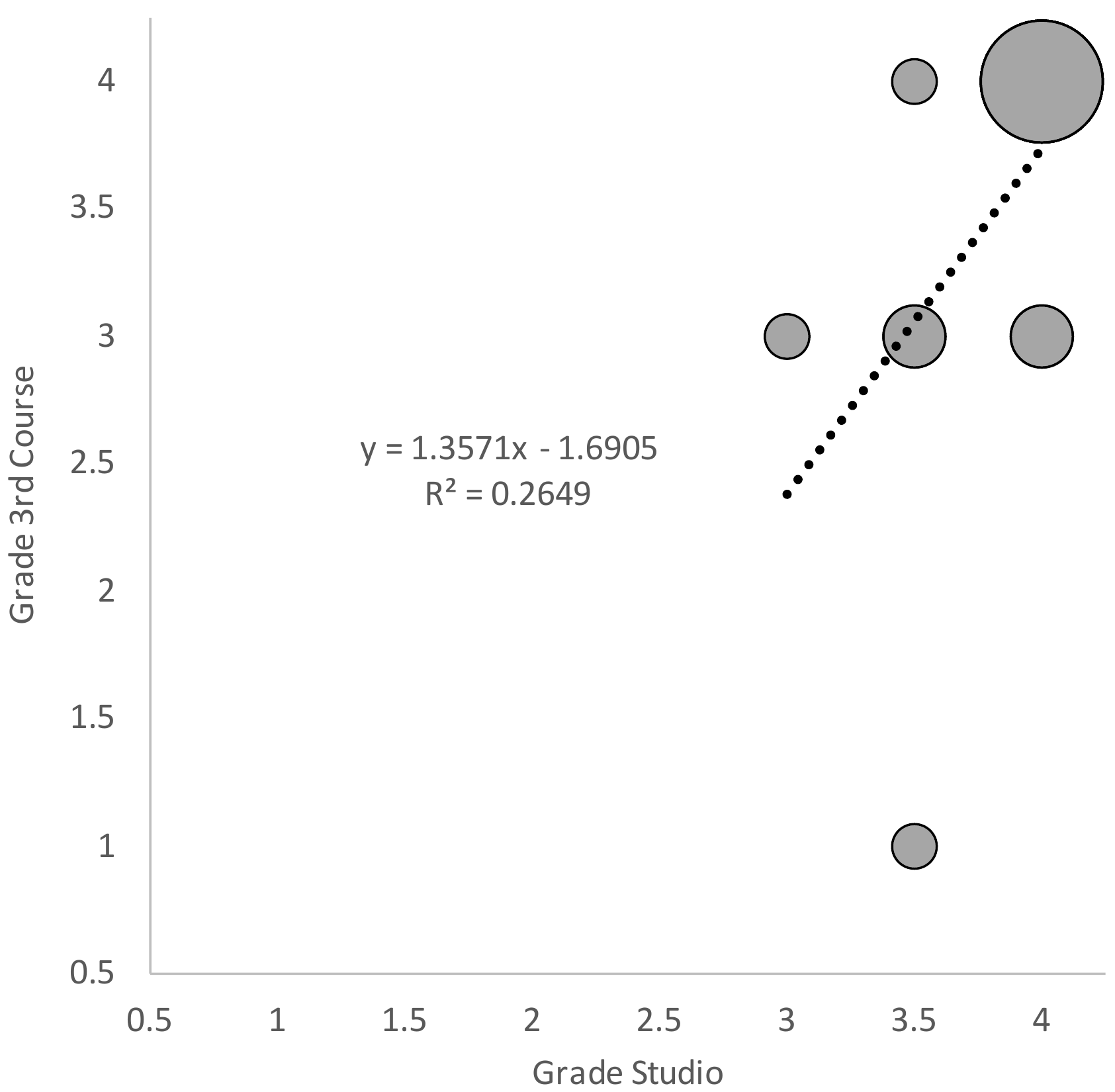}
\end{center}
\caption{Correlations between the grades in different types of introductory courses (left panel \attr{Online}, middle panel \attr{Lecture}, right panel \attr{Studio}) and the subsequent course (\attr{Grade3rd}). The area of the markers is proportional to the number of respondents with the respective combination of grades.\label{fig:gradescor}}
\end{figure*}

Figure~\ref{fig:gradescor} shows the correlation between grades in introductory online, lecture, or studio courses and subsequent third courses. It turns out that the predictive power of the grades in introductory courses is low ($R^2=0.12$ for online, $R^2=0.21$ for lecture, and $R^2=0.32$ for studio courses). Considering only the subsample of students with subsequent advance physics courses (\attr{PhysicsCourse3}), the predictive power is slightly higher for the  online grades, but lower for the  lecture or studio grades ($R^2=0.27$ for online, $R^2=0.09$ for lecture, and $R^2=0.26$ for studio courses).

\section{Discussion}
It was found earlier that having attended introductory physics in a studio setting had generally no significant impact on subsequent grades, compared to lecture settings~\cite{beichner2007}; it is thus not surprising that no significant effect on future grades could be found in this study. However, a grade-focussed view of courses neglects less tangible goals of courses, such as instilling curiosity and interest, as well as increased self-efficacy; we found that studio-based courses are correlated with those desirable attributes, but this of course does not allow for establishing causal relationships. As an example, we cannot make any statements if the more curious students selected the studio courses, if the studio courses instilled that curiosity, or if there is some other, unknown confounding factor. 

As in several previous studies, women were in the minority in the courses under investigation (definitely within the sample, but from observation also within the whole course population), but as opposed to several earlier studies, women outperformed men in terms of grades (not significantly so, but noticeably).
The subsequent courses taken into consideration in this study would be taken by STEM-majors. It is surprising that even in this population, which would arguably have an affinity to science and math and probably identify as a ``STEM-person,'' female students express significantly less interest in physics than their male counterparts. The finding is even more distressing given that women achieved higher physics grades than men, however, it aligns with earlier studies regarding interest in physics~\cite{kalender2019}. In other words, women succeeded in getting better grades than men in spite of professing less interest in the subject, which hints toward a stronger focus solely on grades~\cite{lin}, higher diligence~\cite{kortemeyer09,richardson2013}, or seeing the study of physics simply as a means-to-an-end regarding future studies or career~\cite{barthelemy2020}. It should also be noted that our survey took place after the introductory course sequence was completed, and there are indications that loss of initial interest in physics particularly among women can be the result of these courses~\cite{marshman2018}.

We did not observe the statistically significant preference of women to select studio-based courses found earlier~\cite{shieh2011}, but we did find a preference away from lecture-based courses (of course, this study has a third course mode, online, as a choice, making a significant trend away from one option not necessarily a significant trend toward one particular other one).

Online courses are much more scalable than lecture-based courses, which in turn require significantly less personnel resources than studio-based courses. While a null result technically is no result, the lack of significant findings is consistent with literature~\cite{russell1999no,cavanaugh2015,kortemeyer2023attending,bergeler2021,kortemeyer22hybrid}. Given that we only find very slight differences in success in subsequent courses between the three teaching modes, one has to ask if an allocation of significant additional teaching personnel is justifiable~\cite{chirikov2020}. Arguably, possible differences may have been masked by the fact that students had free choice; they were able to choose the mode that they felt works best for them, which is not a controlled experiment~\cite{kortemeyer2023attending}.

\section{Limitations}
While the survey was distributed to hundreds of students, only 104~students provided valid responses. There may well have been a selection-bias toward the more motivated, engaged students. Also, not finding significant correlations within this sample does not mean that there are none within the full population.

\section{Conclusion}
Overall, we found that the mode of instruction of an introductory, calculus-based physics sequence generally has no significant influence on grades  at the end of a subsequent engineering or more advanced physics course. However, in particular combinations, there was a retained effect on self-efficacy and interest in physics: students who were enrolled in introductory studio-physics courses felt better prepared for the subsequent courses, particularly those that included laboratory or recitation components. Studio classes were also significantly positively correlated with wondering how physics works, while lecture-based courses had a negative correlation with this measure of curiosity about physics.

While female students received slightly but not significantly better grades in all three courses (introductory and subsequent), they generally expressed significantly less interest in physics. In terms of preferred learning scenario, female students who later took advanced physics courses were less likely to select lecture-based introductory physics courses.

None of the non-trivial significant correlations between any of the variables, and none of regression coefficients for grades in future courses involve participation in online courses. This null result opens up questions regarding the best allocation of teaching resources.

\begin{acknowledgments}
We would like to thank the students who participated in this study.
\end{acknowledgments}

\bibliography{ThreeCourses}

\end{document}